# On Uncertainty of Dynamic Systems via State Aggregation Coarse-Graining and State Decomposition Fine-Graining Ways

Lirong Cui[1, 2, *], Xiangchen Li[3], Narayanaswamy Balakrishnan[4]
[1]*Qingdao University, Qingdao 266071, China*
[2]*China Sport Information Center, Beijing 100061, China*
[3]*China Institute of Sport Science, Beijing 100061, China*
[4]*McMaster University, Hamilton, Ontario, Canada L8S 4K1*

## Abstract

Uncertainty is an important feature of dynamic systems, and entropy has been widely used to measure this attribute. In this Letter, we prove that state aggregation and decomposition can decrease and increase the entropy, respectively, of dynamic systems. More than 20 popular entropies in the literature are summarized and analyzed, and it is noted that none of them breaks this property. Finally, pertinent proofs are given for four cases.

*Introduction*

The concept of entropy has a long history. It has been used wisely in many fields in sciences and engineering [1, 2, 3], especially in information theory and quantum theory [4, 5]. A wide array of entropy measures have been discussed in the literature. As an important feature of dynamic systems, uncertainty, which includes chaos, fractal and randomness, has received considerable attention in the literature. One of the difficulties is in capturing the degree of uncertainty in dynamic systems, and entropies are commonly used for this purpose. For dynamic systems, uncertainty variations may result from changes in the number of possible states: one is the state aggregation, in which states are aggregated by widening ranges of variables, reducing the resolution of observations and/or zooming out actions, etc. On the other hand, the number of states could also be changed by state decomposition, in which states are decomposed by reducing ranges of variables, raising the resolution of observations and/or zooming in actions, etc.

All these actions on increasing or decreasing the number of possible states are essentially the same as state aggregation coarse-graining and state decomposition fine-graining ways. Intuitively, entropies may increase as the number of possible states increases and vice versa. In this Letter, we shall establish this for finite discrete distributions in the case of many popular entropies known in the literature.

*Definitions*

Let the set of all $n-$dimensional $(n \geq 2)$ distributions be denoted by

$$\mathcal{P}_n = \{(p_1, p_2, \ldots, p_n) : p_i > 0, \sum_{i=1}^{n} p_i = 1\},$$

$\{1,2,\ldots,n\}$ be the state space of dynamic systems with any partition denoted by $\mathcal{A} = (\mathcal{A}_1, \mathcal{A}_2, \ldots, \mathcal{A}_k)$ $(k \leq n)$, and $\mathbb{R}_0^+ = \{x : x \in [0,\infty)\}$ be the set of nonnegative real numbers. For example, given a specification of probability mass function, $\mathcal{P} \in \mathcal{P}_n$, the corresponding state space of the dynamic system is $\{1,2,\ldots,n\}$, one state aggregation coarse-graining way, denoted by $\mathcal{A}$, is

$$\mathcal{A} = (\mathcal{A}_1, \mathcal{A}_2, \mathcal{A}_3) = (\{1,2\}, \{3,4\}, \{5,\ldots,n\}),$$

where $\mathcal{A}_1 = \{1,2\}, \mathcal{A}_2 = \{3,4\}, \mathcal{A}_3 = \{5,\ldots,n\}$.

For two coarse-graining ways $\mathcal{A}$ and $\mathcal{B}$, if $\mathcal{B}$ can be generated by merging some states in $\mathcal{A}$, then it is expressed as $\mathcal{B} \subseteq \mathcal{A}$. These the entropy is a function $H(\mathcal{P}) : \mathcal{P}_n \to \mathbb{R}_0^+$, where $\mathcal{P} \in \mathcal{P}_n$.

In order to make our result clear, some sets of entropies are defined as follows:

$$S_1 = \{H(\mathcal{P}) : H(\mathcal{P}) = \sum_{i=1}^{n} \varphi(p_i), \varphi(0) = 0, \varphi'(x) \geq \varphi'(x+p),\ 0 \leq x \leq 0.5, 0 \leq p \leq 1-x,$$

$$\text{or } H(\mathcal{P}) = h(\sum_{i=1}^{n} \varphi(p_i)), \varphi(0) = 0, h'(x) > 0 \text{ and } \varphi''(x) < 0 \text{ or } h'(x) < 0 \text{ and } \varphi''(x) > 0\ \}.$$

Note that, if $\varphi(p_i) = \varphi(n, p_i)$, the additional condition that, $\sum_{i=1}^{n} \varphi(n,0) = \text{constant}$, for any $n$, must be satisfied. Also

$$S_2 = \{H(\mathcal{P}) : H(\mathcal{P}) \text{ satisfies Axioms (i)-(iv), and one of A1 to A5}\},$$

where Axiom (i) Positivity: $H(\mathcal{P}) \geq 0$, Axiom (ii) Expandability: $H(\mathcal{P}) = H(p_1, \ldots, p_i, 0, p_{i+1}, \ldots, p_n)$ for any $i \in \{1,\ldots,n\}$, Axiom (iii) Symmetry: $H(\mathcal{P}) = H(p_{\pi(1)}, \ldots,, p_{\pi(n)})$, where $(\pi(1), \ldots, \pi(n))$ is a permutation of $(1,\ldots,n)$, Axiom

(iv) Continuity： $H(\mathcal{P})$ is a continuous function of all variables $p_1,\ldots,p_n$. Further:

A1 [6]: Recursivity, meaning

$$H(p_1,\ldots,p_n)=H(p_1+p_2,p_3,\ldots,p_n)+(p_1+p_2)H\left(\frac{p_1}{p_1+p_2},\frac{p_2}{p_1+p_2}\right);$$

A2 [6]: Separability (or strong additivity), meaning

$$\begin{aligned}&H(p_{11},\ldots,p_{1U},p_{21},\ldots,p_{2U},\ldots,p_{W1},\ldots,p_{WU})\\&=H(p_{1*},p_{2*},\ldots,p_{W*})+\sum_{i=1}^{W}p_{i*}H\left(\frac{p_{i1}}{p_{i*}},\ldots,\frac{p_{iU}}{p_{i*}}\right),\end{aligned}$$

where $p_{i*}=\sum_{j=1}^{U}p_{ij}$;

A3 [7]: Crucial recursivity, meaning

$$H_n(p_1,\ldots,p_{m-1},p_mq_1,\ldots,p_mq_{n-m+1})=H_m(p_1,\ldots,p_m)+p_mH(q_1,\ldots,q_{n-m+1}),$$

where $\sum_{i=1}^{m}p_i=1$ and $\sum_{i=1}^{n-m+1}q_i=1$;

A4 [8]: Let $\mathcal{P}\in\mathcal{P}_n$ and $\mathcal{PQ}=(r_{11},\ldots,r_{nm})\in\mathcal{P}_{nm}$ such that $p_i=\sum_{j=1}^{m}r_{ij}$ and $\mathcal{Q}_{|k}=(q_{1|k},\ldots,q_{m|k})\in\mathcal{P}_m$, where $q_{i|k}=r_{ik}/p_k$ and $\alpha\in\mathbb{R}_0^+$ is a fixed parameter. Then, an entropy $H_{nm}(\mathcal{PQ})$ satisfies

$$H_{nm}(\mathcal{PQ})=H_n(\mathcal{P})+H_m(\mathcal{Q}\,|\,\mathcal{P}),$$

where $H_m(\mathcal{Q}\,|\,\mathcal{P})=f^{-1}(\sum_{k=1}^{n}p_k^{(\alpha)}f(H_m(Q_{|k})))$, $f(x)$ is an invertible continuous function and $\mathcal{P}^{(\alpha)}=(p_1^{(\alpha)},\ldots,p_n^{(\alpha)})\in\mathcal{P}_n$ is an $\alpha$-escort distribution with $p_k^{(\alpha)}=p_k^{\alpha}/\sum_{i=1}^{n}p_i^{\alpha}$.

When $f(x)=x$, then this axiom reduces to the so-called [SK4] (Shannon-Khinchin axioms) [9], i.e.,

$$H_{nm}(\mathcal{PQ})=H_n(\mathcal{P})+\sum_{k=1}^{n}p_k^{(\alpha)}H_m(Q_{|k});$$

A5 [10]: If the entropy satisfies the following postulate, for $\gamma\in\mathbb{R}_0^+$,

$$H_{nm}(\mathcal{PQ})=H_n(\mathcal{P})\oplus_{\gamma}H_m(\mathcal{Q}\,|\,\mathcal{P}),$$

where $H_m(\mathcal{Q}\,|\,\mathcal{P})=f^{-1}(\sum_{k=1}^{n}p_k^{(\alpha)}f(H_m(Q_{|k})))$, the operator $\oplus_{\gamma}$ is defined as

$$u \oplus_\gamma v = u + v + \gamma uv, \quad u, v \in \mathbb{R} .$$

When $f(x) = x$, then this axiom reduces to the so-called [SM4] [11], given by,

$$H_{nm}(\mathcal{PQ}) = H_n(\mathcal{P}) + \sum_{k=1}^{n} p_k^{(\alpha)} H_m(Q_{|k}) + \gamma H_n(\mathcal{P}) \sum_{k=1}^{n} p_k^{(\alpha)} H_m(Q_{|k}),$$

and finally

$$S_3 = \{H(\mathcal{P}) : H(\mathcal{P}) = g(G(\mathcal{P})), G(\mathcal{P}) : \mathcal{P} \to \mathbb{R}_0^+, \text{ is an entropy}, G(\mathcal{P}) \in S_1 \bigcup S_2\},$$

where $g(x)$ is monotonic in $x \in \mathbb{R}_0^+$.

We present a comprehensive list of entropies in Table 1, and the relationships of these entropies to sets $S_1, S_2, S_3$ are given in the table as well.

Table 1 List of entropies to be used in this work

<table>
<tr><th colspan="2">Names of Entropies</th><th>Symbols & Formulas</th><th>Year</th></tr>
<tr><td colspan="2">Shannon entropy [4]</td><td>$H_S(\mathcal{P}) = -\sum_{i=1}^{n} p_i \log p_i$</td><td>1948</td></tr>
<tr><td colspan="2">Rényi entropy [12]</td><td>$R_q(\mathcal{P}) = \frac{1}{1-q} \log\left(\sum_{i=1}^{n} p_i^q\right), \ (0 < q \ \& \ q \neq 1)$</td><td>1961</td></tr>
<tr><td colspan="2">Tsallis entropy [13]</td><td>$T_q(\mathcal{P}) = \frac{1}{1-q}\left(\sum_{i=1}^{n} p_i^q - 1\right), \ (q \geq 0 \ \& \ q \neq 1)$</td><td>1988</td></tr>
<tr><td colspan="2">$(h,\phi)$ -entropy [14]</td><td>$H_{(h,\phi)}(\mathcal{P}) = h\left(\sum_{i=1}^{n} \phi(p_i)\right)$ if $\phi''(x) \leq 0$, $h'(x) \geq 0$, $x \in [0,1]$</td><td>1993</td></tr>
<tr><td rowspan="3">Special cases</td><td>Genetic entropy [15]</td><td>$\phi(x) = x - x^2 - x^2(1-x)^2, \ h(x) = x$</td><td>1973</td></tr>
<tr><td>Paired entropy [14]</td><td>$\phi(x) = -x \log(x) - (1-x)\log(1-x), \ h(x) = x$</td><td>1993</td></tr>
<tr><td>Hypoentropy [16]</td><td>$\phi(x) = \frac{1}{n}\left(1 + \frac{1}{\lambda}\right)\log(1+\lambda) - \frac{1}{\lambda}(1+\lambda x)\log(1+\lambda x)$ , $h(x) = x$</td><td>1980</td></tr>
</table>

| | | | |
|---|---|---|---|
| | $r$ order and $s$ order entropy [17] | $\phi(x)=x^{r}$ , $h(x)=\frac{1}{1-s}\left(x^{\frac{s-1}{r-1}}-1\right)$, $r,s\neq 1, r>0$ | 1975 |
| Universal-group entropy [18] | | $S_U(\mathcal{P})=\sum_{i=1}^{n}p_i G\left(\frac{1}{\ln p_i}\right)$, $G(t)=\sum_{k=0}^{\infty}a_k\frac{t^{k+1}}{k+1}$, $a_0>0$ & $\{a_k\}_{k\in\mathbb{N}}$, $a_k>(k+1)a_{k+1}$, $\forall k\in\mathbb{N}$ | 1975 |
| Special cases | $S_{c,d}$ entropy [19] | $S_{c,d}=\frac{e}{1-c+cd}\sum_{i=1}^{n}\Gamma(1+d,1-c\ln(p_i)-\frac{c}{1-c+cd}$, $\Gamma(a,b)=\int_{b}^{\infty}t^{a-1}e^{-t}\mathrm{d}t$ , $c\in(0,1]$, $d\in\mathbb{R}$, $e$ is the natural constant. | 2011 |
| | $S_\delta$ entropy [20] | $S_\delta=\sum_{i=1}^{n}p_i\left(\ln\frac{1}{p_i}\right)^{\delta}$ , $0<\delta\leq(1+\ln n)$ | 2009 |
| | Borgs -Roditi entropy [21] | $S_{a,b}=\sum_{i=1}^{n}\frac{p_i^b-p_i^a}{a-b}$, $0\leq a,b<1$ | 1998 |
| | Group entropy [22] | $S_G=\frac{1}{\sigma}\sum_{i=1}^{n}\sum_{j=l}^{m}k_j p_j^{-j\sigma}, l,m\in\mathbb{Z}, m-l>0,$ $\sum_{j=l}^{m}k_j=0,\ \sum_{j=l}^{m}jk_j=1, k_m\neq 0, k_l\neq 0$ | 2011 |
| | $S_{III}$ entropy [18] | $S_{III}=\frac{1}{1-q}\sum_{i=1}^{n}p_i[p_i^{2(1-q)}-2p_i^{(1-q)}+p_i^{-(1-q)}]$, $2/3<q<1$ | 2016 |
| | $S_{IV}$ entropy [18] | $S_{IV}=\frac{1}{1-q}\sum_{i=1}^{n}p_i\left[p_i^{-2(1-q)}-\frac{3}{2}p_i^{-(1-q)}+\frac{3}{2}p_i^{(1-q)}-p_i^{2(1-q)}\right]$ | 2016 |
| | Three-parameter entropy [18] | $S_{\alpha,\beta,q}=\frac{1}{1-q}\sum_{i=1}^{n} p_i[\alpha p_i^{-2(1-q)}+\frac{1}{2}(1-3\alpha+\beta)p_i^{-(1-q)}+\frac{1}{2}(\alpha-1-3\beta)p_i^{(1-q)}+\beta p_i^{2(1-q)}],$ $\frac{1}{2}<q<\frac{3}{2},\ 0<\alpha<\frac{1}{2},\ -\frac{1}{4}<\beta<0$ | 2016 |
| Two parameter entropy [23][24] | | $S_{r,k}(\mathcal{P})=-\sum_{i} p_i^{1+r}\left(\frac{p_i^{k}-p_i^{-k}}{2k}\right)$ $(r,k)\in\mathcal{R}=\begin{cases}-\lvert k\rvert\le r\le\lvert k\rvert, & \text{if } 0\le\lvert k\rvert<\frac{1}{2},\\ \lvert k\rvert-1<r<1-\lvert k\rvert, & \text{if } \frac{1}{2}\le\lvert k\rvert<1.\end{cases}$ | 2004 |
| Special cases | Abe-entropy [25] | $\sum_{i=1}^{n}\frac{p_i^{q^{-1}}-p_i^{q}}{q-q^{-1}},\ q=\sqrt{1+k^2}+k,\ r=\sqrt{1+k^2}-1$ | 1997 |
| | Kaniadakis entropy [26] | $\sum_{i=1}^{n} p_i\frac{p_i^{-k}-p_i^{k}}{2k},\ -1<k<1, r=0$ | 2002 |
| | $\gamma$-entropy [27] | $\sum_{i=1}^{n}\frac{p_i^{1-\gamma}-p_i^{1+2\gamma}}{3\gamma},\ r=\frac{1}{2}\gamma, k=\frac{3}{2}\gamma$ | 2005 |
| Nath entropy [28] | | $N(\mathcal{P})=\begin{cases}\tau\sum_{i=1}^{n} p_i\log_2 p_i, & \tau<0,\ \text{for } \lambda=1,\\ \frac{1}{\lambda}\log_2\left(\sum_{i=1}^{n} p_i^{\alpha}\right), & \alpha>0, \lambda(1-\alpha)>0,\ \text{for } \lambda\ne1.\end{cases}$ | 1968 |
| Havrda-Charvát entropy of order $q$ [29] | | $H_q(\mathcal{P})=\frac{\sum_{i=1}^{n} p_i^{q}-1}{2^{1-q}-1}, q\ne1, q>0$ | 1967 |
| Entropy form of order $q$ [30] | | $M_q(\mathcal{P})=\frac{\sum_{i=1}^{n} p_i^{2-q}-1}{q-1}, q\ne1, -\infty<q<2$ | 2006 |
| Additive entropic form of order $q$ [30] | | $M_q^{*}(\mathcal{P})=\frac{\ln\left(\sum_{i=1}^{n} p_i^{2-q}\right)}{q-1}, q\ne1, -\infty<q<2$ | 2006 |

Note: Some special cases are not listed here in this Table.

*Main results*

Entropy, measuring the uncertainty of a dynamic system, can be changed through state aggregation coarse-graining or state decomposition fine-graining ways. The formal result is as given in the following theorem.

**Theorem 1** Given two state aggregation coarse-graining ways $\mathcal{A}=(\mathcal{A}_1,\mathcal{A}_2,\ldots,\mathcal{A}_{k_1})$ and $\mathcal{B}=(\mathcal{B}_1,\mathcal{B}_2,\ldots,\mathcal{B}_{k_2})$, $2\le k_2<k_1\le n$, and $\mathcal{B}\subseteq\mathcal{A}$, if $H(\mathcal{P})\in S_1\bigcup S_2\bigcup S_3$, then

$$H(\mathcal{P}^{\mathcal{B}})\le H(\mathcal{P}^{\mathcal{A}}), \tag{1}$$

where $\mathcal{P}^{\mathcal{A}}=(p_1^{\mathcal{A}},\ldots,p_{k_1}^{\mathcal{A}}), p_i^{\mathcal{A}}=\sum_{j\in\mathcal{A}_i}p_j$ and $\mathcal{P}^{\mathcal{B}}=(p_1^{\mathcal{B}},\ldots,p_{k_2}^{\mathcal{B}}), p_i^{\mathcal{B}}=\sum_{j\in\mathcal{B}_i}p_j$.

**Corollary 1** When $\mathcal{A}=(\mathcal{A}_1,\mathcal{A}_2,\ldots,\mathcal{A}_{k_1})=(\{1\},\{2\},\ldots,\{n\})$ and $k_1=n$, for any entropy $H(\mathcal{P})\in S_1\bigcup S_2\bigcup S_3$, we have the same monotone property: for any partition $\mathcal{B}=(\mathcal{B}_1,\mathcal{B}_2,\ldots,\mathcal{B}_{k_2})$ of $(1,2,\ldots,n)$ $(k_2<n)$, we have

$$H(\mathcal{P}^{\mathcal{B}})\le H(\mathcal{P}). \tag{2}$$

Eq. (1) implies that entropy measuring the uncertainty of a dynamic system can be decreased through state aggregation coarse-graining ways or increased through state decomposition fine-graining ways. It is clear that Eqs. (1) and (2) also hold for mutual entropy (high-dimensional cases) and conditional entropy.

The belonging relations of entropies listed in Table 1 and $S_1,S_2,S_3$ are presented in the following results.

***Relation I***. Some entropies listed in Table 1 belong to set $S_1$, i.e.,

$$\{H_S(\mathcal{P}),T_q(\mathcal{P}),H_{(h,\phi)}(\mathcal{P}),S_U(\mathcal{P}),S_{r,k}(\mathcal{P}),N(\mathcal{P}),H_q(\mathcal{P}),M_q(\mathcal{P})\}\in S_1.$$

***Relation II***. Some entropies listed in Table 1 belong to set $S_2$, i.e.,

$$\{H_S(\mathcal{P}),R_q(\mathcal{P}),T_q(\mathcal{P}),N(\mathcal{P}),H_q(\mathcal{P}),M_q(\mathcal{P})\}\in S_2.$$

***Relation III***. Some entropies listed in Table 1 belong to set $S_3$, i.e.,

$$\{R_q(\mathcal{P}),T_q(\mathcal{P}),H_q(\mathcal{P}),M_q(\mathcal{P}),M_q^*(\mathcal{P})\}\in S_3.$$

*Proofs*

We present the proofs for **Theorem 1** and ***Relations I*** to ***III*** here for some of the entropies, but the overall results are presented in [31], the Supplemental Material.

(i) Let $u(x)=\varphi(x)+\varphi(p)-\varphi(x+p)$; then, $u'(x)=\varphi'(x)-\varphi'(x+p)\geq 0$ because

$$\varphi'(x)\geq\varphi'(x+p),\quad \text{for } 0\leq x\leq 0.5, 0\leq p\leq 1-x.$$

Thus, $u(x)\geq u(0)=0$, i.e., for any probability distribution $\mathcal{P}_n=(p_1,\dots,p_i,\dots,p_j,\dots,p_n)\ (n\geq 2)$, without loss of generality, with the assumption that $p_i\leq p_j$ and $p_i\leq 0.5$, we have

$$u(p_i)=\varphi(p_i)+\varphi(p_j)-\varphi(p_i+p_j)\geq 0, \text{ i.e.,}$$

$$H(\mathcal{P})=\sum_{l=1}^{n}\varphi(p_l)\geq\sum_{l\neq i,j}\varphi(p_l)+\varphi(p_i+p_j)=H(\mathcal{P}_{(i,j)}),$$

where $\mathcal{P}_{(i,j)}=(p_1,\dots,p_{i-1},p_{i+1},\dots,p_{j-1},p_{j+1},\dots,p_n,p_i+p_j)$. It is thus proved that entropies belonging to set $S_1$ satisfy **Theorem 1**.

(ii) Universal-group entropy case: Let $\phi(p)=pG(-\ln p)$. It is clear that $\phi(0)=0$ because the convergent radius is infinity. On the other hand,

$$\phi'(x)=G(-\ln x)-G'(-\ln x),$$

where $G'(t)=\sum_{k=0}^{\infty}a_k t^k$, and so $G'(-\ln x)=\sum_{k=0}^{\infty}(-1)^k a_k(\ln x)^k$. Furthermore,

$$\begin{aligned}\phi'(x)&=G(-\ln x)-G'(-\ln x)\\&=\sum_{k=0}^{\infty}(-1)^{k+1}\frac{a_k}{k+1}(\ln x)^{k+1}-\sum_{k=0}^{\infty}(-1)^k a_k(\ln x)^k\\&=\sum_{k=0}^{\infty}(-1)^{k+1}\frac{a_k}{k+1}(\ln x)^{k+1}-\sum_{k=1}^{\infty}(-1)^k a_k(\ln x)^k-a_0\\&=\sum_{k=0}^{\infty}(-1)^{k+1}\left(\frac{a_k}{k+1}-a_{k+1}\right)(\ln x)^{k+1}-a_0,\end{aligned}$$

and

$$\phi'(x+p)=\sum_{k=0}^{\infty}(-1)^{k+1}\left(\frac{a_k}{k+1}-a_{k+1}\right)[\ln(x+p)]^{k+1}-a_0.$$

Due to the facts that

$$[\ln x]^{2n} \geq [\ln(x+p)]^{2n},\ [\ln x]^{2n-1} \leq [\ln(x+p)]^{2n-1},\ \text{for } 0 \leq x \leq x+p \leq 1,\ \text{ and}$$

$$a_n > (k+1)a_{n+1},\ \ n = 0,1,\ldots,$$

we can obtain

$$\phi'(x) \geq \phi'(x+p),\ (0 \leq x \leq x+p \leq 1).$$

This shows us that the universal-group entropy belongs to set $S_1$ , i.e.,

$$S_U(\mathcal{P}) \in S_1.$$

Thus we have $S_U(\mathcal{P}^{\mathcal{B}}) \leq S_U(\mathcal{P}^{\mathcal{A}})$.

Note that all the operations in the above steps are correct because $G(t)$ is absolutely and uniformly convergent with an infinity radius, which is ensured by the condition $a_k > (k+1)a_{k+1},\ \forall k \in \mathbb{N}$ with $\{a_k\}_{k\in\mathbb{N}} \geq 0,$ where $\mathbb{N} = \{0,1,\ldots\}$.

(iii) The separability (or strong additivity) means

$$\begin{aligned} &H(p_{11},\ldots,p_{1U},p_{21},\ldots,p_{2U},\ldots,p_{W1},\ldots,p_{WU}) \\ &= H(p_{1*},p_{2*},\ldots,p_{W*}) + \sum_{i=1}^{W} p_{i*} H\left(\frac{p_{i1}}{p_{i*}},\ldots,\frac{p_{iU}}{p_{i*}}\right), \end{aligned}$$

where $p_{i*} = \sum_{j=1}^{U} p_{ij}$.

If set $u = 2, w = n-1$ and

$$\begin{cases} p_{11} = p_1 \\ p_{12} = p_2 \end{cases}, \begin{cases} p_{21} = p_3 \\ p_{22} = 0 \end{cases}, \begin{cases} p_{31} = p_4 \\ p_{32} = 0 \end{cases}, \ldots, \begin{cases} p_{w1} = p_n \\ p_{w2} = 0 \end{cases},$$

we obtain

$$\begin{aligned} &H(p_1,p_2,\ldots,p_n) \\ &= H(p_{11},p_{12},p_{21},0,p_{31},0,\ldots,p_{n-1,1},0) \\ &= H(p_{11},p_{12},p_{21},p_{22},\ldots,p_{n-1,1},p_{n-1,2}) \\ &= H(p_1+p_2,p_3,\ldots,p_n) + (p_1+p_2)H\left(\frac{p_1}{p_1+p_2},\frac{p_2}{p_1+p_2}\right) + \sum_{i=2}^{n-1} p_i H(1,0) \\ &= H(p_1+p_2,p_3,\ldots,p_n) + (p_1+p_2)H\left(\frac{p_1}{p_1+p_2},\frac{p_2}{p_1+p_2}\right),\ (n \geq 3). \end{aligned}$$

Thus, we have

$$H(p_{11},p_{12},p_{21},p_{22},\ldots,p_{n-1,1},p_{n-1,2}) \geq H(p_1+p_2,p_3,\ldots,p_n),\ (n \geq 3),\ \text{ i.e.,}$$

$$H(\mathcal{P}^{\mathcal{A}}) \geq H(\mathcal{P}^{\mathcal{B}})\ \text{ if } \mathcal{B} \subset \mathcal{A},$$

providing that entropies satisfying the axiom of separability (belonging to set $S_2$) must follow **Theorem 1**.

(iv) It is known that Tsallis entropy and Rényi entropy satisfies the relationship

$$R_q(\mathcal{P}) = \frac{1}{1-q}\log\{1+(1-q)T_q(\mathcal{P})\}.$$

It is easy to see that the function

$$y = \frac{1}{1-q}\log[1+(1-q)x],\ x \ge 0,\ (0 < q\ \&\ q \ne 1),$$

is an increasing function, and so

$$\begin{aligned} R_q(\mathcal{P}^{\mathcal{A}}) &= \frac{1}{1-q}\log[1+(1-q)T_q(\mathcal{P}^{\mathcal{A}})] \\ &\ge \frac{1}{1-q}\log[1+(1-q)T_q(\mathcal{P}^{\mathcal{B}})] = R_q(\mathcal{P}^{\mathcal{B}}) \end{aligned}$$

if $\mathcal{B} \subset \mathcal{A}$. This is because Tsallis entropy belongs to set $S_1$.

In fact, the following relationships can be given, where the symbols "$\Leftrightarrow$ and $\Updownarrow$" represent the directly relationships existed between two entropies located in the arrows.

$$\begin{array}{ccccc} R_q(\mathcal{P}) & \Leftrightarrow & T_q(\mathcal{P}) & \Leftrightarrow & M_q(\mathcal{P}) \\ & & \Updownarrow & & \Updownarrow \\ & & H_q(\mathcal{P}) & & M_q^*(\mathcal{P}) \end{array}$$

In fact, we can directly prove some entropies follow **Theorem 1**, and we also show it for Shannon entropy.

(v) For two given quantities $0 < p_1 < p_2 < 1$, we have

$$p_1\log\left(1+\frac{p_2}{p_1}\right) \ge 0 \ \text{ and } \ p_2\log\left(1+\frac{p_1}{p_2}\right) \ge 0.$$

Furthermore,

$$p_1\log\left(1+\frac{p_2}{p_1}\right) + p_2\log\left(1+\frac{p_1}{p_2}\right) \ge 0,$$

that is,

$$(p_1+p_2)\log(p_1+p_2) \ge p_1\log(p_1) + p_2\log(p_2),$$

which directly proves that $H(\mathcal{P}^{\mathcal{A}}) \ge H(\mathcal{P}^{\mathcal{B}})$ if $\mathcal{B} \subset \mathcal{A}$.

*Implications*

**Theorem 1** established here yields the following implications.

(i) The entropies are decreasing via state aggregation coarse-graining for dynamic systems. Conversely, the state decomposition fine-graining way for dynamic systems results in increasing the entropies, which is a universal rule for all entropies discussed in the literature. This universal law may be placed as axioms of entropy definitions;

(ii) All the functional related entropies must be consistent with each other, i.e., if there exists the relationship $H_1(\mathcal{P}) = g(H_2(\mathcal{P}))$, then the function $g(x)$ must be nonnegative increasing continuously for $x \in \mathbb{R}_0^+$; otherwise, the two entropies would provide contradictory message on the system uncertainty;

(iii) The dynamic system equipped with an entropy must satisfy the inequality

$$H(\mathcal{P}) \geq H(p_1 + p_2, p_3, \ldots, p_n) \text{ or } H(\mathcal{P}) \geq H(\mathcal{P}_{(i,j)}),$$

and **Theorem 1** can be easily extended to high-dimensional discrete distributions and the conditional entropy cases.

*Discussion & Conclusion*

The main result $H(\mathcal{P}^{\mathcal{A}}) \geq H(\mathcal{P}^{\mathcal{B}})$ is in accordance with one's intuition, i.e., the decomposition of one state in a dynamic system increases the system uncertainty, and conversely, the aggregation of two or more states in a dynamic system decreases the system uncertainty. The feature requirements of set $S_1$ can be reduced to $\varphi''(x) \leq 0$ because it can result in $\varphi'(x) \geq \varphi'(x+p)$ for $\forall x \in [0, 0.5]$ and $0 \leq x + p \leq 1$. In fact, if $\varphi''(x) \leq 0$ does not hold, then the entropy $H(\mathcal{P}) = \sum_{i=1}^{n} \varphi(p_i)$ may not satisfy (1) in general. Here is an example to demonstrate this. Let an entropy $H_E(\mathcal{P}) := \sum_{i=1}^{n} \varphi_E(p_i),\ n > 2,$ be such that

$$\varphi_E(x) = \begin{cases} x, & 0 \leq x \leq 0.25, \\ 2x - 0.25, & 0.25 \leq x < 0.5, \\ 1.75 - 2x, & 0.5 \leq x < 0.75, \\ 1 - x, & 0.75 \leq x \leq 1, \end{cases}$$

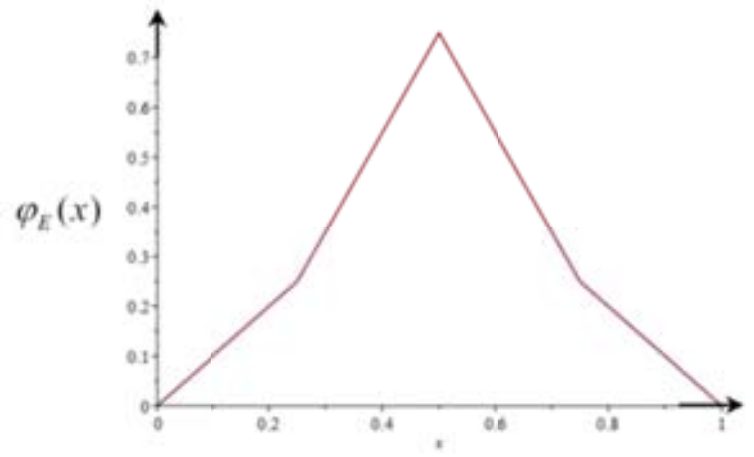

Figure 1. The curve of $\varphi_E(x)$

whose curve is presented in Figure 1. When $\mathcal{P}=(p_1,p_2,p_3)=(0.2,0.3,0.5)$, then

$$H_E(\mathcal{P})=H_E(p_1,p_2,p_3)=H_E(0.2,0.3,0.5)=1.3, \text{ and}$$

$H_E(p_1+p_2,p_3)=H_E(0.5,0.5)=1.5$. We then have

$$H_E(0.2,0.3,0.5)=1.3<H_E(0.5,0.5)=1.5 .$$

In fact, this entropy satisfies positivity, expandability, symmetry and continuity, but the function $\varphi_E(x)$ breaks the condition

$$1=\frac{\mathrm{d}\varphi_E(x)}{\mathrm{d}x}\geq\frac{\mathrm{d}\varphi_E(x+p)}{\mathrm{d}x}=2, \quad \text{for } 0<x<0.25,\ 0.25<x+p\leq 0.5.$$

Furthermore, we have

$$H_E(0.25,0.25,0.25,0.25)=1<H_E(0.2,0.25,0.25,0.3)=1.05,$$

which breaks the maximal entropy principle when $p_i=1/n$.

The axiom of separability or recursivity is as important feature of entropy, which may be further extended. If there exists a relation between two entropies, then this relation must be an increasing function, which would maintain the consistence of the two entropies. Thus, if one of two entropies possesses (1), then the other must satisfies (1) as well.

We have established here that $H(\mathcal{P}^{\mathcal{A}})\geq H(\mathcal{P}^{\mathcal{B}})$ for many known entropies, which implies that state aggregation and decomposition can, respectively, decrease and increase the entropy. So far, we have not found any known entropy to break the result presented here. The corresponding proofs are given here, but for all the cases, the proofs can be found in [31] in the Supplemental Material. The results can be used not only in describing the changes of uncertainty for dynamic systems, but also in finding the bounds of entropies in information, quantum and other areas.

This work was supported by (i) National Key R&D Program of China (2020YFC2007201), and (ii) National Natural Science Foundation of China (71871021).

# Supplementary information:
# On Uncertainty of Dynamic Systems via State Aggregation Coarse-Graining and State Decomposition Fine-Graining Ways

Lirong Cui[1, 2, *], Xiangchen Li[3], Narayanaswamy Balakrishnan[4]
[1]*Qingdao University, Qingdao 266071, China*
[2]*China Sport Information Center, Beijing 100061, China*
[3]*China Institute of Sport Science, Beijing 100061, China*
[4]*McMaster University, Hamilton, Ontario, Canada L8S 4K1*

**Contents**

## 1. Definitions & Preliminaries

Let the set of all $n-$ dimensional $(n \geq 2)$ distributions be denoted by

$$\mathcal{P}_n = \{(p_1, p_2, \ldots, p_n) : p_i > 0, \sum_{i=1}^{n} p_i = 1\}$$

and the state space of dynamic systems be $\{1, 2, \ldots, n\}$, its any partition denoted by $\mathcal{A} = (\mathcal{A}_1, \mathcal{A}_2, \ldots, \mathcal{A}_k)$ $(k \leq n)$, $\mathbb{R}_0^+ = \{x : x \in [0, \infty)\}$ be the set of nonnegative real numbers. For example, given a specifications of probability mass function, $\mathcal{P} \in \mathcal{P}_n$, the corresponding state space of the dynamic system is $\{1, 2, \ldots, n\}$, one state aggregation coarse-graining way, denoted by $\mathcal{A}$, is

$$\mathcal{A} = (\mathcal{A}_1, \mathcal{A}_2, \mathcal{A}_3) = (\{1, 2\}, \{3, 4\}, \{5, \ldots, n\}),$$

where $\mathcal{A}_1 = \{1, 2\}, \mathcal{A}_2 = \{3, 4\}, \mathcal{A}_3 = \{5, \ldots, n\}$.

For two coarse-graining ways $\mathcal{A}$ and $\mathcal{B}$, if $\mathcal{B}$ can be generated by merging some states in $\mathcal{A}$, then it is expressed as $\mathcal{B} \subseteq \mathcal{A}$. These the entropy is a function $H(\mathcal{P}): \mathcal{P}_n \to \mathbb{R}_0^+$, where $\mathcal{P} \in \mathcal{P}_n$.

In order to make our result clear, some sets of entropies are defined as follows:

$$S_1 = \{H(\mathcal{P}): H(\mathcal{P}) = \sum_{i=1}^{n} \varphi(p_i), \varphi(0) = 0, \varphi'(x) \geq \varphi'(x+p),\ 0 \leq x \leq 0.5, 0 \leq p \leq 1-x,$$

$$\text{or } H(\mathcal{P}) = h(\sum_{i=1}^{n} \varphi(p_i)), \varphi(0) = 0, h'(x) > 0 \text{ and } \varphi''(x) < 0 \text{ or } h'(x) < 0 \text{ and } \varphi''(x) > 0\},$$

Note that, if $\varphi(p_i) = \varphi(n, p_i)$, the additional condition that, $\sum_{i=1}^{n} \varphi(n, 0) = \text{constant}$, for any $n$, must be satisfied. Also

$$S_2 = \{H(\mathcal{P}): H(\mathcal{P}) \text{ satisfies Axioms (i) -(iv), and one of A1 to A5}\},$$

where Axiom (i) Positivity: $H(\mathcal{P}) \geq 0$, Axiom (ii) Expandability: $H(\mathcal{P}) = H(p_1, \ldots, p_i, 0, p_{i+1}, \ldots, p_n)$ for any $i \in \{1, \ldots, n\}$, Axiom (iii) Symmetry: $H(\mathcal{P}) = H(p_{\pi(1)}, \ldots,, p_{\pi(n)})$, where $(\pi(1), \ldots, \pi(n))$ is a permutation of $(1, \ldots, n)$, Axiom (iv) Continuity: $H(\mathcal{P})$ is a continuous function of all variables $p_1, \ldots, p_n$.

A1: Recursivity, meaning

$$H(p_1, \ldots, p_n) = H(p_1 + p_2, p_3, \ldots, p_n) + (p_1 + p_2) H\left(\frac{p_1}{p_1 + p_2}, \frac{p_2}{p_1 + p_2}\right);$$

A2: Separability (or strong additivity), meaning

$$\begin{aligned} &H(p_{11}, \ldots, p_{1U}, p_{21}, \ldots, p_{2U}, \ldots, p_{W1}, \ldots, p_{WU}) \\ &= H(p_{1*}, p_{2*}, \ldots, p_{W*}) + \sum_{i=1}^{W} p_{i*} H\left(\frac{p_{i1}}{p_{i*}}, \ldots, \frac{p_{iU}}{p_{i*}}\right), \end{aligned}$$

where $p_{i*} = \sum_{j=1}^{U} p_{ij}$;

A3: Crucial recursivity, meaning

$$\begin{aligned} &H_n(p_1, \ldots, p_{m-1}, p_m q_1, \ldots, p_m q_{n-m+1}) \\ &= H_m(p_1, \ldots, p_m) + p_m H(q_1, \ldots, q_{n-m+1}), \end{aligned}$$

where $\sum_{i=1}^{m} p_i = 1$ and $\sum_{i=1}^{n-m+1} q_i = 1$;

A4: Let $\mathcal{P} \in \mathcal{P}_n$ and $\mathcal{PQ} = (r_{11}, \ldots, r_{nm}) \in \mathcal{P}_{nm}$ such that $p_i = \sum_{j=1}^{m} r_{ij}$ and

$\mathcal{Q}_{|k} = (q_{1|k}, \dots, q_{m|k}) \in \mathcal{P}_m$, where $q_{i|k} = r_{ik} / p_k$ and $\alpha \in \mathbb{R}_0^+$ is a fixed parameter. Then, an entropy $H_{nm}(\mathcal{PQ})$ satisfies

$$H_{nm}(\mathcal{PQ}) = H_n(\mathcal{P}) + H_m(\mathcal{Q} \mid \mathcal{P}),$$

where $H_m(\mathcal{Q} \mid \mathcal{P}) = f^{-1}(\sum_{k=1}^{n} p_k^{(\alpha)} f(H_m(Q_{|k})))$, where $f$ is an invertible continuous function and $\alpha$ -escort distribution $\mathcal{P}^{(\alpha)} = (p_1^{(\alpha)}, \dots, p_n^{(\alpha)}) \in \mathcal{P}_n$, where $p_k^{(\alpha)} = p_k^{\alpha} / \sum_{i=1}^{n} p_i^{\alpha}$.

When $f(x) = x$, then this axiom reduces to the so-called [SK4] (Shannon-Khinchin axioms), i.e.,

$$H_{nm}(\mathcal{PQ}) = H_n(\mathcal{P}) + \sum_{k=1}^{n} p_k^{(\alpha)} H_m(Q_{|k});$$

A5: If the entropy satisfies the following postulate, for $\gamma \in \mathbb{R}_0^+$,

$$H_{nm}(\mathcal{PQ}) = H_n(\mathcal{P}) \oplus_{\gamma} H_m(\mathcal{Q} \mid \mathcal{P}),$$

where $H_m(\mathcal{Q} \mid \mathcal{P}) = f^{-1}(\sum_{k=1}^{n} p_k^{(\alpha)} f(H_m(Q_{|k})))$, where $f$ is an invertible continuous function, and the operator $\oplus_{\gamma}$ is defined as

$$u \oplus_{\gamma} v = u + v + \gamma uv, \quad u, v \in \mathbb{R}.$$

When $f(x) = x$, then this axiom reduces to

$$H_{nm}(\mathcal{PQ}) = H_n(\mathcal{P}) + \sum_{k=1}^{n} p_k^{(\alpha)} H_m(Q_{|k}) + \gamma H_n(\mathcal{P}) \sum_{k=1}^{n} p_k^{(\alpha)} H_m(Q_{|k}),$$

and finally

$S_3 = \{H(\mathcal{P}) : H(\mathcal{P}) = f(G(\mathcal{P})), G(\mathcal{P}) : \mathcal{P} \to \mathbb{R}_0^+,$ is another entropy, $G(\mathcal{P}) \in S_1 \bigcup S_2\}$.

## 2. Entropies to be used in the Letter

All entropies used in the Letter are listed in Table 1 as below.

Table 1 List of entropies to be used in the Letter

| Names of Entropies | Symbols & Formulas | Year |
|---|---|---|

| | | | |
|---|---|---|---|
| Shannon entropy [4] | | $H_S(\mathcal{P}) = -\sum_{i=1}^{n} p_i \log p_i$ | 1948 |
| Rényi entropy [12] | | $R_q(\mathcal{P}) = \frac{1}{1-q}\log\left(\sum_{i=1}^{n} p_i^q\right),\ (0<q \ \&\ q\neq 1)$ | 1961 |
| Tsallis entropy [13] | | $T_q(\mathcal{P}) = \frac{1}{1-q}\left(\sum_{i=1}^{n} p_i^q - 1\right),\ (q\geq 0 \ \&\ q\neq 1)$ | 1988 |
| $(h,\phi)$ -entropy [14] | | $H_{(h,\phi)}(\mathcal{P}) = h\left(\sum_{i=1}^{n}\phi(p_i)\right)$ if $\phi''(x)\leq 0$, $h'(x)\geq 0$, or $\phi''(x)\geq 0$, $h'(x)<0$, $x\in[0,1]$ | 1993 |
| Special cases | Genetic entropy [15] | $\phi(x) = x - x^2 - x^2(1-x)^2$, $h(x)=x$ | 1973 |
| | Paired entropy [14] | $\phi(x) = -x\log(x)-(1-x)\log(1-x)$, $h(x)=x$ | 1993 |
| | Hypoentropy [16] | $\phi(x) = \frac{1}{n}\left(1+\frac{1}{\lambda}\right)\log(1+\lambda) - \frac{1}{\lambda}(1+\lambda x)\log(1+\lambda x)$ , $h(x)=x$ | 1980 |
| | $r$ order and $s$ order entropy [17] | $\phi(x)=x^r$, $h(x) = \frac{1}{1-s}\left(x^{\frac{s-1}{r-1}}-1\right)$, $r,s\neq 1, r>0$ | 1975 |
| Universal-group entropy [18] | | $S_U(\mathcal{P}) = \sum_{i=1}^{n} p_i G\left(\frac{1}{\ln p_i}\right)$, $G(t) = \sum_{k=0}^{\infty} a_k \frac{t^{k+1}}{k+1}$, $a_0>0$ & $\{a_k\}_{k\in\mathbb{N}}$, $a_k>(k+1)a_{k+1}$, $\forall k\in\mathbb{N}$ | 1975 |
| Special cases | $S_{c,d}$ entropy [19] | $S_{c,d} = \frac{e}{1-c+cd}\sum_{i=1}^{n}\Gamma(1+d, 1-c\ln(p_i)) - \frac{c}{1-c+cd}$, $\Gamma(a,b) = \int_b^{\infty} t^{a-1}e^{-t}\mathrm{d}t$, $c\in(0,1]$, $d\in\mathbb{R}$, $e$ is the natural constant. | 2011 |
| | $S_\delta$ entropy | $S_\delta = \sum_{i=1}^{n} p_i\left(\ln\frac{1}{p_i}\right)^{\delta}$ , $0<\delta\leq(1+\ln n)$ | 2009 |
| | [20] | | |
| | Borgs -Roditi entropy [21] | $S_{a,b}=\sum_{i=1}^{n}\frac{p_i^b-p_i^a}{a-b},\ 0\le a,b<1$ | 1998 |
| | Group entropy [22] | $S_G=\frac{1}{\sigma}\sum_{i=1}^{n}\sum_{j=l}^{m}k_j p_j^{-j\sigma}, l,m\in\mathbb{Z}, m-l>0,$ $\sum_{j=l}^{m}k_j=0,\ \sum_{j=l}^{m}jk_j=1, k_m\ne 0, k_l\ne 0$ | 2011 |
| | $S_{III}$ entropy [18] | $S_{III}=\frac{1}{1-q}\sum_{i=1}^{n}p_i[p_i^{2(1-q)}-2p_i^{(1-q)}+p_i^{-(1-q)}],$ $2/3<q<1$ | 2016 |
| | $S_{IV}$ entropy [18] | $S_{IV}=\frac{1}{1-q}\sum_{i=1}^{n}p_i[p_i^{-2(1-q)}-\frac{3}{2}p_i^{-(1-q)}+\frac{3}{2}p_i^{(1-q)}-p_i^{2(1-q)}]$ | 2016 |
| | Three -parameter entropy [18] | $S_{\alpha,\beta,q}=\frac{1}{1-q}\sum_{i=1}^{n}p_i[\alpha p_i^{-2(1-q)}+\frac{1}{2}(1-3\alpha+\beta)p_i^{-(1-q)}$ $+\frac{1}{2}(\alpha-1-3\beta)p_i^{(1-q)}+\beta p_i^{2(1-q)}],$ $\frac{1}{2}<q<\frac{3}{2},\ 0<\alpha<\frac{1}{2},\ -\frac{1}{4}<\beta<0$ | 2016 |
| Two parameter entropy [23][24] | | $S_{r,k}(\mathcal{P})=-\sum_i p_i^{1+r}\left(\frac{p_i^k-p_i^{-k}}{2k}\right)$ $(r,k)\in\mathcal{R}=\begin{cases}-\lvert k\rvert\le r\le\lvert k\rvert, & \text{if } 0\le\lvert k\rvert<\frac{1}{2},\\ \lvert k\rvert-1<r<1-\lvert k\rvert, & \text{if } \frac{1}{2}\le\lvert k\rvert<1.\end{cases}$ | 2004 |
| Special cases | Abe-entropy [25] | $\sum_{i=1}^{n}\frac{p_i^{q^{-1}}-p_i^q}{q-q^{-1}},\ q=\sqrt{1+k^2}+k,\ r=\sqrt{1+k^2}-1$ | 1997 |
| | Kaniadakis entropy [26] | $\sum_{i=1}^{n}p_i\frac{p_i^{-k}-p_i^k}{2k},\ -1<k<1, r=0$ | 2002 |
| | $\gamma$ -entropy [27] | $\sum_{i=1}^{n}\frac{p_i^{1-\gamma}-p_i^{1+2\gamma}}{3\gamma},\ r=\frac{1}{2}\gamma, k=\frac{3}{2}\gamma$ | 2005 |
| Nath entropy [28] | | $N(\mathcal{P})=\begin{cases}\tau\sum_{i=1}^{n}p_i\log_2 p_i, & \tau<0,\ \text{for } \lambda=1,\\ \frac{1}{\lambda}\log_2\left(\sum_{i=1}^{n}p_i^{\alpha}\right), & \alpha>0, \lambda(1-\alpha)>0,\ \text{for } \lambda\neq 1.\end{cases}$ | 1968 |
| Havrda-Charvát entropy of order $q$ [29] | | $H_q(\mathcal{P})=\frac{\sum_{i=1}^{n}p_i^{q}-1}{2^{1-q}-1}, q\neq 1, q>0$ | 1967 |
| Entropy form of order $q$ [30] | | $M_q(\mathcal{P})=\frac{\sum_{i=1}^{n}p_i^{2-q}-1}{q-1}, q\neq 1, -\infty<q<2$ | 2006 |
| Additive entropic form of order $q$ [30] | | $M_q^{*}(\mathcal{P})=\frac{\ln\left(\sum_{i=1}^{n}p_i^{2-q}\right)}{q-1}, q\neq 1, -\infty<q<2$ | 2006 |

Note: Some special cases are not listed here in this Table.

## 3. Proof for Theorem 1

### 3.1. The case of $S_1$

We shall prove that: If $H(\mathcal{P})\in S_1$, then $H(\mathcal{P}^{\mathcal{B}})\leq H(\mathcal{P}^{\mathcal{A}})$.

*Proof.* Let $u(x)=\varphi(x)+\varphi(p)-\varphi(x+p)$, then $u'(x)=\varphi'(x)-\varphi'(x+p)\geq 0$ because

$$\varphi'(x)\geq\varphi'(x+p),\ \text{ for } 0\leq x\leq 0.5, 0\leq p\leq 1-x.$$

Thus, it has $u(x)\geq u(0)=0$, i.e., for any probability distribution $\mathcal{P}_n=(p_1,\dots,p_i,\dots,p_j,\dots,p_n)$ $(n\geq 2)$, without loss of generality, with the assumption that $p_i\leq p_j$ and $p_i\leq 0.5$, we have

$$u(p_i)=\varphi(p_i)+\varphi(p_j)-\varphi(p_i+p_j)\geq 0\text{, i.e.,}$$

$$H(\mathcal{P})=\sum_{l=1}^{n}\varphi(p_l)\geq\sum_{l\neq i,j}\varphi(p_l)+\varphi(p_i+p_j)=H(\mathcal{P}_{(i,j)}),$$

where $\mathcal{P}_{(i,j)} = (p_1, \ldots, p_{i-1}, p_{i+1}, \ldots, p_{j-1}, p_{j+1}, \ldots, p_n, p_i + p_j)$. It is proved that entropies belonging to set $S_1$ satisfy **Theorem 1**.

### 3.2. The case of $S_2$

We shall prove that: If $H(\mathcal{P}) \in S_2$, then $H(\mathcal{P}^{\mathcal{B}}) \le H(\mathcal{P}^{\mathcal{A}})$.

(1). For the case of recursivity, it means

$$H(p_1, \ldots, p_n) = H(p_1 + p_2, p_3, \ldots, p_n) + (p_1 + p_2) H\left(\frac{p_1}{p_1 + p_2}, \frac{p_2}{p_1 + p_2}\right),$$

which is clearly to result in $H(\mathcal{P}^{\mathcal{B}}) \le H(\mathcal{P}^{\mathcal{A}})$.

(2). For the separability (or strong additivity) case, it means

$$\begin{aligned} & H(p_{11}, \ldots, p_{1U}, p_{21}, \ldots, p_{2U}, \ldots, p_{W1}, \ldots, p_{WU}) \\ & = H(p_{1*}, p_{2*}, \ldots, p_{W*}) + \sum_{i=1}^{W} p_{i*} H\left(\frac{p_{i1}}{p_{i*}}, \ldots, \frac{p_{iU}}{p_{i*}}\right), \end{aligned}$$

where $p_{i*} = \sum_{j=1}^{U} p_{ij}$.

If set $u = 2, w = n - 1$ and

$$\begin{cases} p_{11} = p_1 \\ p_{12} = p_2 \end{cases}, \begin{cases} p_{21} = p_3 \\ p_{22} = 0 \end{cases}, \begin{cases} p_{31} = p_4 \\ p_{32} = 0 \end{cases}, \ldots, \begin{cases} p_{w1} = p_n \\ p_{w2} = 0 \end{cases},$$

we obtain

$$\begin{aligned} & H(p_1, p_2, \ldots, p_n) \\ & = H(p_{11}, p_{12}, p_{21}, 0, p_{31}, 0, \ldots, p_{n-1,1}, 0) \\ & = H(p_{11}, p_{12}, p_{21}, p_{22}, \ldots, p_{n-1,1}, p_{n-1,2}) \\ & = H(p_1 + p_2, p_3, \ldots, p_n) + (p_1 + p_2) H\left(\frac{p_1}{p_1 + p_2}, \frac{p_2}{p_1 + p_2}\right) + \sum_{i=2}^{n-1} p_i H(1, 0) \\ & = H(p_1 + p_2, p_3, \ldots, p_n) + (p_1 + p_2) H\left(\frac{p_1}{p_1 + p_2}, \frac{p_2}{p_1 + p_2}\right), \quad (n \ge 3). \end{aligned}$$

Thus, it can conclude that

$$H(p_{11}, p_{12}, p_{21}, p_{22}, \ldots, p_{n-1,1}, p_{n-1,2}) \ge H(p_1 + p_2, p_3, \ldots, p_n), \quad (n \ge 3), \text{ i.e.,}$$

$$H(\mathcal{P}^{\mathcal{A}}) \ge H(\mathcal{P}^{\mathcal{B}}) \text{ if } \mathcal{B} \subset \mathcal{A}.$$

It is proved that entropies satisfying the axiom of separability (belonging to set $S_2$) must follow **Theorem 1**.

(3). For the case of the crucial recursivity, it means

$$H_n(p_1,\dots,p_{m-1},p_mq_1,\dots,p_mq_{n-m+1}) = H_m(p_1,\dots,p_m) + p_mH(q_1,\dots,q_{n-m+1}),$$

where $\sum_{i=1}^{m} p_i = 1$ and $\sum_{i=1}^{n-m+1} q_i = 1$.

If we let $n-m=1$, then it has

$$\begin{aligned} & H_{m+1}(p_1,\dots,p_{m-1},p_mq_1,p_mq_2) \\ &= H_{m+1}(p_1,\dots,p_{m-1},p_m^{(1)},p_m^{(2)}) \\ &= H_m(p_1,\dots,p_{m-1},p_m^{(1)}+p_m^{(2)}) + p_mH_2(q_1,q_2), \end{aligned}$$

where $p_m^{(1)} = p_mq_1, p_m^{(2)} = p_m(1-q_1)$, which implies that

$$H_{m+1}(p_1,\dots,p_{m-1},p_m^{(1)},p_m^{(2)}) \ge H_m(p_1,\dots,p_{m-1},p_m^{(1)}+p_m^{(2)}).$$

Thus it can continue the procedure successfully, this is because the following fact holds true. The set of equations,

$$\begin{cases} p_i = xy, \\ p_j = (1-x)y, \end{cases} \text{ has a solution } \begin{cases} x = \dfrac{p_i}{p_i+p_j}, \\ y = p_i + p_j, \end{cases} \text{ for any } 0 < p_i + p_j \le 1,\ p_i, p_j > 0.$$

(4). For case (4), let $\mathcal{P} \in \mathcal{P}_n$ and $\mathcal{PQ} = (r_{11},\dots,r_{nm}) \in \mathcal{P}_{nm}$ such that $p_i = \sum_{j=1}^{m} r_{ij}$ and $\mathcal{Q}_{|k} = (q_{1|k},\dots,q_{m|k}) \in \mathcal{P}_m$, where $q_{i|k} = r_{ik}/p_k$ and $\alpha \in \mathbb{R}_0^+$ is a fixed parameter. Then an entropy $H_{nm}(\mathcal{PQ})$ satisfies

$$H_{nm}(\mathcal{PQ}) = H_n(\mathcal{P}) + H_m(\mathcal{Q}\,|\,\mathcal{P}),$$

where $H_m(\mathcal{Q}\,|\,\mathcal{P}) = f^{-1}(\sum_{k=1}^{n} p_k^{(\alpha)} f(H_m(Q_{|k})))$, where $f$ is an invertible continuous function and $\alpha$ -escort distribution $\mathcal{P}^{(\alpha)} = (p_1^{(\alpha)},\dots,p_n^{(\alpha)}) \in \mathcal{P}_n$, where $p_k^{(\alpha)} = p_k^{\alpha} / \sum_{i=1}^{n} p_i^{\alpha}$.

If set $m=2$, $\sum_{i=1}^{n} p_i = 1$, and

$$\begin{cases} r_{11} = p_1 \\ r_{12} = p_2 \end{cases}, \begin{cases} r_{21} = p_3 \\ r_{22} = 0 \end{cases}, \begin{cases} r_{31} = p_4 \\ r_{32} = 0 \end{cases}, \dots, \begin{cases} r_{n1} = p_n \\ r_{n2} = 0 \end{cases},$$

it has

$$\mathcal{Q}_{|1} = \left( \frac{p_1}{p_1 + p_2}, \frac{p_2}{p_1 + p_2} \right) \in \mathcal{P}_2, \quad \mathcal{Q}_{|k} = \left( \frac{p_k}{p_k}, \frac{0}{p_k} \right) = (1,0) \in \mathcal{P}_2, (k = 2,3,\ldots,n).$$

Thus it obtains

$$H_{nm}(\mathcal{PQ}) \geq H_n(\mathcal{P}) \Leftrightarrow H(p_1, p_2, \ldots, p_n) \geq H(p_1 + p_2, p_3, \ldots, p_n),$$

This is because $H_m(\mathcal{Q} \mid \mathcal{P}) = f^{-1}(\sum_{k=1}^{n} p_k^{(\alpha)} f(H_m(Q_{|k}))) \geq 0$.

(5). For case (5). If an entropy satisfies the following postulate, for $\gamma \in \mathbb{R}_0^+$,

$$H_{nm}(\mathcal{PQ}) = H_n(\mathcal{P}) \oplus_\gamma H_m(\mathcal{Q} \mid \mathcal{P}),$$

where $H_m(\mathcal{Q} \mid \mathcal{P}) = f^{-1}(\sum_{k=1}^{n} p_k^{(\alpha)} f(H_m(Q_{|k})))$, where $f$ is an invertible continuous function, and the operator $\oplus_\gamma$ is defined as

$$u \oplus_\gamma v = u + v + \gamma uv, \quad u, v \in \mathbb{R}.$$

When $f(x) = x$, then this axiom reduces to

$$H_{nm}(\mathcal{PQ}) = H_n(\mathcal{P}) + \sum_{k=1}^{n} p_k^{(\alpha)} H_m(Q_{|k}) + \gamma H_n(\mathcal{P}) \sum_{k=1}^{n} p_k^{(\alpha)} H_m(Q_{|k}).$$

The proof is similar to that in case (4), because $H_m(\mathcal{Q} \mid \mathcal{P}) \geq 0$ and $\gamma \geq 0$.

### 3.3. The case of $S_3$

Because of the definition of set $S_3$ and the increasing property of function $g(x)$, if one entropy

$$G(\mathcal{P}) : \mathcal{P} \to \mathbb{R}_0^+, \quad G(\mathcal{P}) \in S_1 \cup S_2,$$

then

$$H(\mathcal{P}^{\mathcal{A}}) = g(G(\mathcal{P}^{\mathcal{A}})) \geq g(G(\mathcal{P}^{\mathcal{B}})) = H(\mathcal{P}^{\mathcal{B}}).$$

## 4. Proofs for Relations I to III

***Relation I***. Some entropies listed in Table 1 belong to set $S_1$, i.e.,

$$\{H_S(\mathcal{P}), T_q(\mathcal{P}), H_{(h,\phi)}(\mathcal{P}), S_U(\mathcal{P}), S_{r,k}(\mathcal{P}), N(\mathcal{P}), H_q(\mathcal{P}), M_q(\mathcal{P}), M_q^*(\mathcal{P})\} \in S_1.$$

(1) For the case of $H_S(\mathcal{P})$: Shannon entropy.

Since $\varphi(x) = -x\log(x)$ for Shannon entropy case, it is easy to know $\varphi''(x) < 0$.

(2) For the case of $T_q(\mathcal{P})$: Tsallis entropy.

Since $\varphi(x) = \dfrac{x^q - x}{1-q}$ for Tsallis entropy case, $\varphi''(x) <= -qx^{q-2} < 0$.

(3) For the case of $(h,\phi)$ -entropy.

Since $\phi''(x) < 0$, thus $\phi(x+p) \le \phi(x) + \phi(p),\ 0 < x \le 0.5,\ 0 < x+p \le 1,$ further, it has

$$h(\sum_{i=1}^{n}\phi(p_i)) \ge h(\phi(p_1+p_2) + \sum_{i=3}^{n}\phi(p_i)).$$

(4) For the case of $S_U(\mathcal{P})$: Universal entropy.

Let $\phi(p) = pG(-\ln p)$. It is clear to know $\phi(0) = 0$ because the convergent radius is infinity. On the other hand, it has

$$\phi'(x) = G(-\ln x) - G'(-\ln x),$$

where $G'(t) = \sum_{k=0}^{\infty} a_k t^k$, and so $G'(-\ln x) = \sum_{k=0}^{\infty}(-1)^k a_k (\ln x)^k$. Furthermore,

$$\begin{aligned}
\phi'(x) &= G(-\ln x) - G'(-\ln x) \\
&= \sum_{k=0}^{\infty}(-1)^{k+1}\frac{a_k}{k+1}(\ln x)^{k+1} - \sum_{k=0}^{\infty}(-1)^k a_k(\ln x)^k \\
&= \sum_{k=0}^{\infty}(-1)^{k+1}\frac{a_k}{k+1}(\ln x)^{k+1} - \sum_{k=1}^{\infty}(-1)^k a_k(\ln x)^k - a_0 \\
&= \sum_{k=0}^{\infty}(-1)^{k+1}[\frac{a_k}{k+1} - a_{k+1}](\ln x)^{k+1} - a_0,
\end{aligned}$$

and

$$\phi'(x+p) = \sum_{k=0}^{\infty}(-1)^{k+1}[\frac{a_k}{k+1} - a_{k+1}][\ln(x+p)]^{k+1} - a_0.$$

Due to the fact that

$$[\ln x]^{2n} \ge [\ln(x+p)]^{2n},\ [\ln x]^{2n-1} \le [\ln(x+p)]^{2n-1},\ \text{for } 0 \le x \le x+p \le 1, \text{ and}$$

$$a_n > (k+1)a_{n+1},\ \ n = 0,1,\dots,$$

we can obtain

$$\phi'(x) \ge \phi'(x+p),\ (0 \le x \le x+p \le 1).$$

This shows us that the universal-group entropy belongs to set $S_1$, i.e.,

$$S_U(\mathcal{P}) \in S_1.$$

Thus, we have $S_U(\mathcal{P}^{\mathcal{B}}) \le S_U(\mathcal{P}^{\mathcal{A}})$.

(5) For the case of $S_{r,k}(\mathcal{P})$ : Two-parameter entropy.

In terms of the results in [27], we know that

$\ln_{\{k,r\}}(p) := p^r \dfrac{p^k - p^{-k}}{2k}$ is an increasing concave function in the region of parameter space, then we can obtain

$$\frac{\mathrm{d}^2}{\mathrm{d}x^2}[x\ln_{\{k,r\}}(x)] \le 0, \quad \text{for } (k,r) \in \mathcal{R}. \quad \mathcal{R} = \begin{cases} -|k| \le r \le |k|, & \text{if } 0 \le |k| < \dfrac{1}{2}, \\ |k|-1 < r < 1-|k|, & \text{if } \dfrac{1}{2} \le |k| < 1. \end{cases}$$

It tells us that the function $x\ln_{\{k,r\}}(x)$ is a concave function for $x \in [0,1]$ and $(k,r) \in \mathcal{R}$. (see [27])

For the case of $N(\mathcal{P})$: Nath entropy.

(i) For $\tau < 0$ and $\lambda = 1$, it is clear to know $N_{\tau<0,\lambda=1}(\mathcal{P}) \in S_1$.

(ii) For $\alpha > 0, \lambda(1-\alpha) > 0$ and $\lambda \ne 1$, let $h(x) = \dfrac{1}{\lambda}\log_2(x)$ and $\phi(x) = x^{\alpha}$, we can know that $h'(x) > 0$ and $\phi''(x) < 0$ if $\lambda > 0$, otherwise, $h'(x) < 0$ and $\phi''(x) > 0$. Thus it is proved $N(\mathcal{P}) \in S_1$.

(6) For the case of $H_q(\mathcal{P})$ : Havrda-Charvát entropy of order $q$.

Let $\varphi(x) = \dfrac{x^q - x}{2^{1-q} - 1}$, we have $\varphi''(x) = \dfrac{1}{2^{1-q} - 1}[q(q-1)x^{q-2}] < 0,$ for $q \ne 1, q > 0$, then it has $H_q(\mathcal{P}) \in S_1$.

(8). For the case of $M_q(\mathcal{P})$ : Entropy form of order $q$.

Let $\varphi(x) = \frac{x^{2-q} - x}{q-1}$, we have $\varphi''(x) = (q-2)x^{-q} < 0$, for $q \neq 1, -\infty < q < 2$. Thus it has $M_q(\mathcal{P}) \in S_1$.

(9). For the case of $M_q^*(\mathcal{P})$: Additive entropic form of order $q$.

Let $h(x) = \frac{\ln(x)}{q-1}$, $\phi(x) = x^{2-q}$. Similarly, we have

$$h'(x) > 0 \text{ and } \phi''(x) < 0, \text{ for } 1<q<2, \text{ and } h'(x) < 0 \text{ and } \phi''(x) > 0, \text{ for } -\infty<q<1.$$

***Relation II***. Some entropies listed in Table 1 belong to set $S_2$, i.e.,

$$\{H_S(\mathcal{P}), R_q(\mathcal{P}), T_q(\mathcal{P}), N(\mathcal{P}), H_q(\mathcal{P}), M_q(\mathcal{P})\} \in S_2.$$

(1). Shannon entropy:

See Shannon-Khinchin axiom SA4 in Reference [31].

(2). Rényi entropy:

See the generalized additivity: NSK4 in reference [33]

(3). Tsallis entropy:

See reference [10]

(4). Nath entropy:

See reference [8]

(5). Havrda-Charvát entropy of order $q$:

See reference [10]

(6). Entropy form of order $q$:

Because the entropy form of order $q$ is a special case of Tsallis entropy, thus see reference [10].

***Relation III***. Some entropies listed in Table 1 belong to set $S_3$, i.e.,

$$\{R_q(\mathcal{P}), T_q(\mathcal{P}), H_q(\mathcal{P}), M_q(\mathcal{P}), M_q^*(\mathcal{P})\} \in S_3.$$

Proof. Since the following relationships can be obtained in terms of the definitions of related entropies, which are shown in Figure 1.

$$R_q(\mathcal{P}) \Leftrightarrow T_q(\mathcal{P}) \Leftrightarrow M_q(\mathcal{P})$$

$$\Updownarrow \qquad \Updownarrow$$

$$H_q(\mathcal{P}) \quad M_q^*(\mathcal{P})$$

Figure 1. The relationships among 5 entropies. The legend "$\Leftrightarrow$ and $\Updownarrow$" represent the directly relationships existed between two entropies located in the arrows. For example, $R_q(\mathcal{P}) \Leftrightarrow T_q(\mathcal{P})$, it has $R_q(\mathcal{P}) = (1-q)^{-1}\log[1+(1-q)T_q(\mathcal{P})]$. Other relationships can be given easily, here omitted them. Based on **Relations I** and **II**, the proof is completed.

## 5. Some directive proofs for Theorem 1

### (1) Shannon entropy:

For two quantities $0 < p_1 < p_2 < 1$ such that $p_1 + p_2 \le 1$, then it has

$$-p_1 \log p_1 - p_2 \log p_2 \ge -(p_1 + p_2)\log(p_1 + p_2).$$

*Proof.* For two given quantities $0 < p_1 < p_2 < 1$, it has

$$p_1 \log(1 + \frac{p_2}{p_1}) \ge 0 \quad \text{and} \quad p_2 \log(1 + \frac{p_1}{p_2}) \ge 0.$$

Furthermore, they can result in that

$$p_1 \log(1 + \frac{p_2}{p_1}) + p_2 \log(1 + \frac{p_1}{p_2}) \ge 0,$$

that is,

$$(p_1 + p_2)\log(p_1 + p_2) \ge p_1 \log(p_1) + p_2 \log(p_2),$$

which proves the result.

### (2) Rényi entropy:

Definition of Rényi's entropy is

$$R_q(\boldsymbol{p}) = \frac{1}{1-q}\log_2(\sum_{i=1}^{n} p_i^q), \qquad (0 < q).$$

For the case of $0 < q < 1$, since $a^x$ $(a > 0, x > 0)$ is an increasing convex function, then it has,

$$\frac{p_1^q + p_2^q}{2} \ge (\frac{p_1 + p_2}{2})^q \Leftrightarrow p_1^q + p_2^q \ge 2^{1-q}(p_1 + p_2)^q \ge (p_1 + p_2)^q.$$

Thus, it has

$$R_q(p_1, p_2, \ldots, p_n) \ge R_q(p_1 + p_2, p_3, \ldots, p_n) \quad \text{for} \quad 0 < q < 1.$$

For the case of $q > 1$, for any integer $m \geq 2$, it has

$$p_1^m + p_2^m \leq (p_1 + p_2)^m .$$

Let $q = m + r,\ 0 \leq r < 1$, then it has

$$(\frac{p_1}{p_1 + p_2})^r p_1^m + (\frac{p_2}{p_1 + p_2})^r p_2^m \leq (p_1 + p_2)^m ,$$

because

$$0 < \max\{(\frac{p_1}{p_1 + p_2})^r, (\frac{p_2}{p_1 + p_2})^r\} < 1 .$$

Then it obtains $p_1^q + p_2^q \leq (p_1 + p_2)^q$ for $q > 1$.

Furthermore, it has

$$R_q(p_1, p_2, \ldots, p_n) \geq R_q(p_1 + p_2, p_3, \ldots, p_n) \quad \text{for } q > 1 .$$

**(3) Tsallis entropy:**

Because Tsallis entropy is

$$T_q(\boldsymbol{p}) = \frac{1}{1-q}\sum_{i=1}^{n} p_i^q - 1 .$$

Similarly, let $u(x) = x^q + p_2^q - (x + p_2)^q$, then can prove that $u(p_1) < 0$ if $q > 1$ and $u(p_1) > 0$ if $0 < q < 1$.

**(4) $\phi$-entropies:**

Burbea and Rao [34] studied a very important family of entropies, called $\phi$-entropies, whose general expression is

$$\phi\text{-entropy},\quad H_\phi(\boldsymbol{p}) = \sum_{i=1}^{n} \phi(p_i) .$$

Particularizing for different $\phi(x)$ functions:

Shannon Entropy: $\phi(x) = -x\log(x)$

Genetic entropy [15]: $\phi(x) = x - x^2 - x^2(1-x)^2$

Paired entropy: $\phi(x) = -x\log(x) - (1-x)\log(1-x)$

Hypoentropy [16]: $\phi(x)=\frac{1}{n}(1+\frac{1}{\lambda})\log(1+\lambda)-\frac{1}{\lambda}(1+\lambda x)\log(1+\lambda x)$

For the **paired entropy**, let $\varphi(x)=-(1-x)\log(1-x)$, then it will prove that

$$\varphi(p_1)+\varphi(p_2)\geq\varphi(p_1+p_2)\quad\text{for}\quad 0<p_1\leq p_2<1\quad\text{and}\quad 0<p_1+p_2\leq 1.$$

*Proof.* First let

$$\begin{aligned}v(x)&=-[\varphi(x)+\varphi(p_2)-\varphi(x+p_2)]\\&=(1-x)\log(1-x)+(1-p_2)\log(1-p_2)-(1-x-p_2)\log(1-x-p_2),\ \ 0<x\leq p_2.\end{aligned}$$

$$\frac{dv(x)}{dx}=\log(1-x-p_2)-\log(1-x)<0,\ \ x\leq 1-p_2,$$

i.e., the function $v(x)$ decreases in $x$. Thus it has $v(x)\leq v(0)=0$, which proves our result

$$\varphi(x)+\varphi(p_2)\geq\varphi(x+p_2),\quad\text{for } x\in(0,1-p_2).$$

For **genetic entropy,** it is easy to prove that

$$\phi(x)+\phi(p_2)-\phi(x+p_2)\geq 0,\quad 0<x<p_2,$$

where $\phi(x)=x-x^2-x^2(1-x)^2$, which tells us that the genetic entropy satisfies the coarse-graining way decreasing the entropy.

For Hypoentropy: let $\phi(x)=\frac{1}{n}(1+\frac{1}{\lambda})\log(1+\lambda)-\frac{1}{\lambda}(1+\lambda x)\log(1+\lambda x)$,

It can also be proved that

$$\phi(x)+\phi(p_2)-\phi(x+p_2),\quad 0<x<p_2,\ \lambda>0.$$

All results presented above tell us that Theorem 1 holds.